\documentstyle[pra,aps,eqsecnum]{revtex}

\newcommand{\ba}{\begin{eqnarray}}
\newcommand{\ea}{\end{eqnarray}}

\begin{document}

\title{Robustness of Decoherence-Free Subspaces for Quantum Computation}
\author{D. Bacon$^{(1,2)}$, D.A. Lidar$^{(2)}$, and K.B. Whaley$^{(2)}$}
\address{Department of Physics$^{(1)}$ and Department of Chemistry$^{(2)}$,\\
The University of California, Berkeley, CA 94720 }
\date{\today}
\maketitle

\begin{abstract}
It was shown recently [D.A. Lidar {\it et al.}, Phys. Rev. Lett. {\bf 81},
2594 (1998)] that within the framework of the semigroup Markovian master
equation, decoherence-free (DF) subspaces exist which are stable to first
order in time to a perturbation. Here this result is extended to the
non-Markovian regime and generalized. In particular, it is shown that within
both the semigroup and the non-Markovian operator sum representation, DF
subspaces are stable to all orders in time to a symmetry-breaking
perturbation. DF subspaces are thus ideal for quantum memory applications.
For quantum computation, however, the stability result does not extend beyond the first order.  Thus, to perform robust quantum computation in DF
subspaces, they must be supplemented with quantum error correcting codes.
\end{abstract}

\section{Introduction}

The power promised by quantum computers \cite{Steane:98} has initiated an
intense scrutiny of the physical viability of these computers \cite
{Unruh:95,Haroche:96,Landauer:96}. The central obstacle in the experimental
realization of such computers has proven to be maintaining the quantum
coherence of states which form the cornerstone of the speedup promised by
quantum computers. The main cause of this degradation of the quantum
coherence is the coupling of the computer to the environment, and the
subsequent decoherence induced by this coupling. To overcome this
difficulty, {\em Quantum Error Correcting Codes} (QECC) inspired by
classical coding theory have been developed \cite{Bennett:98review}. These
codes are ``active'' in the sense that decoherence is fought by continuous
application of error correction procedures to quantum bits (qubits), which
are encoded over the Hilbert space of several physical qubits. Another
approach has emerged more recently, in which the structure of the physical
decoherence process is used to protect the precious quantum coherence \cite
{Duan:98,Zanardi:97a,Zanardi:97c,Zanardi:98a,Lidar:PRL98,Lidar:98QECC-DFS}.
These ``passive'' error {\em prevention} codes rely on symmetries of the
decoherence process to encode qubits into states which reside in the {\em 
decoherence-free} (DF) {\em subspaces} of multiple physical qubit systems.
The conditions under which such DF subspaces can exist have been established
in both the Lindblad (Markovian) formulation \cite{Zanardi:98a,Lidar:PRL98}
and for the non-Markovian case \cite{Zanardi:97a}.

The non-Markovian formulation of the reduced dynamics of a subsystem is used
extensively in the quantum computation literature, where it is known as the
``operator sum representation'' (OSR) \cite{Kraus:83}. While it is {\em 
exact }, it is not always clear how to separate system from bath in this
approach, since for any finite-dimensional bath the dynamics are {\em 
reversible} (albeit potentially with a very large Poincar\'{e} recurrence
time) \cite{vanWonderen:95} as long as no measurements are made. In the
Lindblad formulation, on the other hand, the dynamics being described are 
{\it a priori } that of the system alone, subject to preservation of
complete positivity of the system density matrix. The dynamics are
irreversible, but the price paid is that the Markovian approximation must be
invoked \cite{Lindblad:76}. The first aim of this paper is to clarify the
relation between these two formulations. We do this with a derivation of the
Lindblad semigroup master equation (SME) from the OSR in a way which
highlights the differences and similarities between the two. Derivations of
master equations are numerous, starting with the work of Zwanzig and others
in the late 1950s \cite{Zwanzig:61}: our current derivation is novel in that
it starts from a fixed-basis form of the OSR which is constructed to
formally resemble the SME, yet is still exact. This fixed-basis
representation allows one to clearly identify the manner in which the temporal
coarse-graining assumption is invoked in the process of making the
transition from non-Markovian to Markovian dynamics, as well as the
consequences of this. The fixed-basis OSR equation leads to several
important results concerning the decay of quantum coherence. First, we show
that any finite total Hamiltonian will have a zero first order decoherence
rate in the non-Markovian case, but that this feature can be destroyed by
the coarse-grained time averaging made upon going to the Markovian limit.
Second, a non-vanishing first order decay rate within the non-Markovian OSR
formulation necessarily implies a singularity of the total Hamiltonian. We
show explicitly how this accounts for the behavior seen in the prototypical
example of phase damping, which is commonly used in the quantum computation
literature.

The second, and main, focus of this paper is the use of the fixed-basis OSR
representation to provide a general stability analysis of DF subspaces. In
Ref.~\cite{Lidar:PRL98} it was shown that, within the SME, DF subspaces are
stable to symmetry-breaking perturbations, to first order in time. This
leads to a lowering of the threshold for fault-tolerant quantum computation 
\cite{Knill:98,Aharonov:96} in DF subspaces. This stability result was
obtained for the ``memory-fidelity'', i.e., for a quantum computer {\em not}
subject to external ``programming pulses''. Since DF subspaces also appear
in the non-Markovian setting \cite{Zanardi:97a,Lidar:98QECC-DFS}, it is
natural to enquire then whether the robustness of DF subspaces with respect
to perturbations extends also to the non-Markovian situation. We make a
stability analysis within the OSR here to address this, and find that the
robustness of the DF subspaces is upheld in the more general situation.
Indeed, we show that in fact, within both the SME and the OSR, DF subspaces
are stable to a symmetry-breaking perturbation to {\em all} orders of time.

The generality of this stability result for memory fidelity does not extend
however to the ``dynamical fidelity'', which measures the preservation of
quantum coherence in the case of a quantum computer that {\em is} subject to
external programming pulses. We examine this dynamical fidelity here and
find that in this case, a DF subspace is stable only to first order in time.
Thus while a DF subspace can drastically extend the decoherence time for
quantum memory, operations performed on the DF subspace must be performed
rapidly (in comparison to the perturbing error rate) in order achieve a
similar extension of the decoherence time for implementing quantum
computation. Barring methods that rely on symmetrizing operations on a
time-scale faster than the primary decoherence rate \cite{Zanardi:99a}, DF
subspaces were, however, never meant to be a complete solution to the
problem of decoherence on a quantum computer. Their usefulness lies in the
elimination of the primary source of decoherence and the subsequent
lengthening of the decoherence time to one determined solely by the
decoherence due to perturbing errors. DF subspaces should be supplemented by
a QECC in order to achieve a decoherence free quantum computer. This is
possible using a concatenation of DF and QECC codes, as was shown in Ref.~ 
\cite{Lidar:98QECC-DFS}. The instability of DF subspaces while the system is
evolving thus sets a lower bound on how rapidly operations on the DF
subspace must be performed in order realize the robustness of DF subspaces.

The structure of the paper is as follows. In Sec.~\ref{OSR} a brief review
of the OSR formalism is presented, followed by the derivation of the fixed
basis form for the OSR equation in Sec.~\ref{newOSR}. This equation is used
to derive the SME in Sec.~\ref{OSR-SG}. In Sec.~\ref{short} we turn to the
main subject of the paper, namely short-time expansions of the fidelity.
Addressing first the fidelity over the entire system Hilbert space, we
derive the first order decoherence rate within the OSR and show that it
vanishes generally in the OSR for non-Markovian dynamics, provided that the
total Hamiltonian is of a non-singular form. This condition precludes the
situation of a system coupled to an infinite number of degrees of freedom.
We then show that this first order rate may become finite as a result of the
coarse-grained time averaging performed on going to the Markovian limit. The
generality of the non-Markovian result appears initially surprising, since
there exist elementary examples of non-vanishing first order decoherence
rates in the non-Markovian situation. In Sec.~\ref{PD}, we show with the
example of the well-known case of decoherence due to phase damping, how this
reflects an underlying singularity in the total Hamiltonian. Sec.~\ref{SPB}
then deals with the special issue of stability for the DF subspaces. After a
brief summary of the conditions for DF subspaces in the two approaches, we
then show that DF subspaces have enhanced stability over the general system
Hilbert space, namely that they are stable to {\em all} orders of
symmetry-breaking perturbations, both within the non-Markovian OSR and the
Markovian (SME) limit. In Sec.~\ref{dyna-F}, we then address the ``dynamical
fidelity'' of DF subspaces under external fields, corresponding to a
``quantum computer program.'' We show that this is stable to a lesser
extent, possessing a vanishing {\em first order} decoherence rate in both
the SME and the OSR, but having non-zero terms of higher order in time. We
conclude with a summary and discussion of the implications for quantum
computation in Section~\ref{conc}.

\section{The Operator Sum Representation}

\subsection{Brief Review}

\label{OSR}

The dynamics of a quantum system $S$ coupled to a bath $B$, which together
form a closed system, evolves unitarily under the combined system-bath
Hamiltonian ${\bf H}_{SB}={\bf H\otimes I}_{B}+{\bf I}_{S}\otimes {\bf H}
_{B}+{\bf H}_{I}$. Here ${\bf H}$, ${\bf H}_{B}$ and ${\bf H}_{I}$ are,
respectively, the system, bath and interaction Hamiltonians, and ${\bf I}$
is the identity operator. Assuming that $S$ and $B$ are initially decoupled,
so that the total initial density matrix is a tensor product of the system
and bath density matrices ($\rho $ and $\rho _{B}$ respectively), the system
dynamics are described by the reduced density matrix:

\begin{equation}
\rho (0)\longmapsto \rho (t)={\rm Tr}_{B}[{\bf U}(\rho \otimes \rho _{B}) 
{\bf U}^{\dagger }].  \label{eq:dyna}
\end{equation}
Here Tr$_{B}$ is the partial trace over the bath and ${\bf U}=\exp (-\frac{i 
}{\hbar }{\bf H}_{SB}t)$. By using a spectral decomposition for the bath, $
\rho _{B}=\sum_{\nu }\nu |\nu \rangle \langle \nu |$, and introducing a
basis $\{|n\rangle \}_{n=1}^{N}$ for the $N$-dimensional system Hilbert
space ${\cal H}$, this can be rewritten in the OSR as \cite
{Kraus:83,Schumacher:96a}:

\begin{equation}
\rho (t)=\sum_{i=0}^{K}{\bf A}_{i}(t)\,\rho (0)\,{\bf A}_{i}^{\dagger }(t),
\label{eq:OSR}
\end{equation}
where the Kraus operators $\{{\bf A}_{i}\}$ have matrix elements given by:

\begin{equation}
\lbrack {\bf A}_{i}]_{mn}(t)=\sqrt{\nu }\langle m|\langle \mu |{\bf U}
(t)|\nu \rangle |n\rangle \;;\qquad i=(\mu ,\nu ).  \label{eq:Amunu}
\end{equation}
Here $|\nu,\mu \rangle$ are basis elements of the bath Hilbert space,
and $K=N_B^2$, where $N_B$ is the number of bath degrees of freedom. Also, by
unitarity of ${\bf U}$, one derives the normalization condition

\begin{equation}
\sum_{i=0}^K {\bf A}_i^\dagger {\bf A}_i={\bf I} ,  \label{eq:OSRnorm}
\end{equation}
which guarantees preservation of the trace of $\rho$:

\begin{equation}
{\rm Tr}[\rho (t)]={\rm Tr}[\sum_{i}{\bf A}_{i}\,\rho (0)\,{\bf A}
_{i}^{\dagger }]={\rm Tr}[\rho (0)\sum_{i}{\bf A}_{i}^{\dagger }{\bf A}
_{i}]= {\rm Tr}[\rho (0)].
\end{equation}
The Kraus operators belong to the Hilbert-Schmidt space ${\cal A}({\cal H})$
(itself a Hilbert space) of bounded operators acting on the system Hilbert
space, and are represented by $N\times N$ matrices, just like $\rho $. $
{\cal A}({\cal H})$ is endowed with the scalar product 
\begin{equation}
\langle {\bf A}_{i},{\bf A}_{j}\rangle ={\rm Tr}[{\bf A}_{i}{\bf A}
_{j}^{\dagger }].  \label{eq:IP}
\end{equation}

\subsection{Fixed-Basis Form of the Operator Sum Representation}

\label{newOSR}

While the OSR evolution equation, Eq.~(\ref{eq:OSR}), is perfectly
general, it presents difficulties when trying to separate out the
unitary evolution of the system from the possibly non-unitary
decoherence which occurs from the coupling of the system to the
bath. The reason is that in general, each Kraus operator will contain
a contribution from both the unitary and the non-unitary components of
the evolution. When one makes the assumption of Markovian dynamics, however, we shall see that the semigroup master
equation (SME) does separate the evolution of the system into unitary
and non-unitary parts \cite{Alicki:87}. This motivates us to
manipulate the OSR into a form similar to the SME, but without making
any Markovian assumption.

It is convenient for this purpose to introduce a {\em fixed} operator basis
for ${\cal A}({\cal H})$ \cite{Chuang:97c}. Let $\{{\bf K}_{\alpha
}\}_{\alpha =0}^{M}$, with ${\bf K}_{0}={\bf I}$, be such a basis, so that
the expansion of the Kraus operators is given by:

\begin{equation}
{\bf A}_{i}(t)=\sum_{\alpha =0}^{M}b_{i\alpha }(t){\bf K}_{\alpha }.
\label{eq:A-F}
\end{equation}
For example, $\{{\bf K}_{\alpha }\}_{\alpha =1}^{M}$ could be the generators
of the Lie algebra $su(N)$ ($M=N^{2}-1$), or some sub-algebra thereof (with $
M<N^2-1)$ \cite{Lidar:PRL98} . Under this expansion, the OSR evolution
equation, Eq.~(\ref{eq:OSR}), becomes

\begin{equation}
\rho (t) = \sum_{\alpha ,\beta =0}^{M}\chi _{\alpha \beta }(t){\bf K}
_{\alpha }\rho (0){\bf K}_{\beta }^{\dagger },  \label{eq:chiOSR}
\end{equation}
where $\chi_{\alpha \beta}(t)$ is the hermitian matrix

\begin{equation}
\chi _{\alpha \beta }(t)=\sum_{i=0}^{K}b_{i\alpha }(t)b_{i\beta }^{\ast }(t).
\label{eq:defchi}
\end{equation}
Likewise the normalization condition, Eq.~(\ref{eq:OSRnorm}), is given by,

\begin{equation}
\sum_{\alpha ,\beta =0}^{M}\chi _{\alpha \beta }(t){\bf K}_{\beta }^{\dagger
}{\bf K}_{\alpha }={\bf I}.  \label{eq:chiOSRnorm}
\end{equation}
Next we separate out the action of the identity on both Eq.~(\ref{eq:chiOSR}) 
and Eq.~(\ref{eq:chiOSRnorm}) yielding:

\begin{equation}
\rho (t)=\chi _{00}\rho (0)+\sum_{\alpha =1}^{M}\left[ \chi _{\alpha 0}(t) 
{\bf K}_{\alpha }\rho (0)+\chi _{0\alpha }(t)\rho (0){\bf K}_{\alpha
}^{\dagger }\right] +\sum_{\alpha ,\beta =1}^{M}\chi _{\alpha \beta }(t){\bf 
K}_{\alpha }\rho (0){\bf K}_{\beta }^{\dagger },  \label{eq:chiOSRnt}
\end{equation}

\begin{equation}
\chi _{00}{\bf I}+\sum_{\alpha =1}^{M}\left[ \chi _{0\alpha }(t){\bf K}
_{\alpha }^{\dagger }+\chi _{\alpha 0}(t){\bf K}_{\alpha }\right]
+\sum_{\alpha ,\beta =1}^{M}\chi _{\alpha \beta }(t) {\bf K} _{\beta
}^{\dagger } {\bf K}_{\alpha } ={\bf I}.  \label{eq:chiOSRnormnt}
\end{equation}
Multiplying Eq.~(\ref{eq:chiOSRnormnt}) by ${\frac{1}{2}}\rho (0)$
separately from both the left and the right, adding the resulting equations,
and substituting the resulting expression for $\chi_{00}\rho (0)$ into
Eq.~(\ref{eq:chiOSRnt}), we find: 

\begin{equation}
\rho(t)-\rho(0) = - {\frac{i }{\hbar}} [ {\bf S}(t),\rho(0)] + {\frac{1 }{2}}
\sum_{\alpha,\beta=1}^M \chi_{\alpha \beta}(t) \left( [{\bf K}_\alpha,
\rho(0) {\bf K}_\beta^\dagger] + [{\bf K}_\alpha \rho(0),{\bf K}
_\beta^\dagger] \right),  \label{eq:newOSR}
\end{equation}
where ${\bf S}(t)$ is the hermitian operator defined by

\begin{equation}
{\bf S}(t)={\frac{i\hbar }{2}}\sum_{\alpha =1}^{M}\left[ \chi _{\alpha 0}(t)
{\bf K}_{\alpha }-\chi _{0\alpha }(t){\bf K}_{\alpha }^{\dagger }\right] .
\label{eq:defJ}
\end{equation}
Eq.~(\ref{eq:newOSR}) is the desired result: it represents a fixed-basis OSR
evolution equation. This generally resembles the SME in form, but this
resemblance should be considered with caution: it can be shown by explicit
resummation that for purely {\em unitary} evolution, $\rho (t)={\bf U}
(t)\rho (0){\bf U}^{\dagger }(t)$, the terms in the fixed-basis OSR are $[
{\bf S}(t),\rho (0)]=\left[ \sin \left( \frac{{\bf H}t}{\hbar }\right) ,\rho
(0)\right] $ and ${\frac{1}{2}}\sum_{\alpha \beta }\left( \cdots \right) =
{\bf U}(t)\rho (0){\bf U}^{\dagger }(t)+(i/\hbar )\left[ \sin \left( \frac{
{\bf H}t}{\hbar }\right) ,\rho (0)\right] -\rho (0)$. Hence the first term
alone $[{\bf S}(t),\rho (0)]$ does not necessarily account for the entire
unitary dynamics, as one might naively be led to suspect.

\subsection{Comparison of the Fixed-Basis Operator Sum Representation
Equation with the Semigroup Master Equation}

\label{OSR-SG}

We recall that in the semigroup approach, under the assumptions of (i)
Markovian dynamics, (ii) complete positivity, and (iii) initial decoupling
between the system and the bath, the system evolves according to the SME 
\cite{Lindblad:76,Alicki:87,Pechukas:94}:

\begin{eqnarray}
\frac{\partial \rho (t)}{\partial t} &=&{\tt L}[\rho (t)]\equiv -\frac{i}{
\hbar }[{\bf H},\rho (t)]+{\tt L}_{D}[\rho (t)]  \label{eq:SG1} \\
{\tt L}_{D}[\rho (t)] &=&\frac{1}{2}\sum_{\alpha ,\beta =1}^{M}a_{\alpha
\beta }([{\bf F}_{\alpha },\rho (t){\bf F}_{\beta }^{\dagger }]+[{\bf F}
_{\alpha }\rho (t),{\bf F}_{\beta }^{\dagger }]),  \label{eq:SG}
\end{eqnarray}
where $a_{\alpha \beta }$ is a constant hermitian matrix. This equation
bears a clear resemblance to Eq.~(\ref{eq:newOSR}). In fact, taking the
derivative of Eq.~(\ref{eq:newOSR}), we find

\begin{equation}
\frac{\partial \rho (t)}{\partial t}=-{\frac{i}{\hbar }}[\dot{{\bf S}}
(t),\rho (0)]+{\frac{1}{2}}\sum_{\alpha ,\beta =1}^{M}\dot{\chi}_{\alpha
\beta }(t)\left( [{\bf K}_{\alpha },\rho (0){\bf K}_{\beta }^{\dagger }]+[ 
{\bf K}_{\alpha }\rho (0),{\bf K}_{\beta }^{\dagger }]\right) .
\label{eq:newOSR2}
\end{equation}
Noticing the subtle and important differences between the SME Eq.~(\ref
{eq:SG}) and this OSR evolution equation (\ref{eq:newOSR2}) allows us to
understand the exact manner in which the semigroup evolution can
arises from the OSR evolution under the above mentioned
conditions.  The most important difference between these two equations
is the fact that the SME provides a prescription for determining $\rho
(t)$ at all times $t$, given $\rho(t^{\prime})$ at any other time
$t^{\prime }\geq 0$, whereas Eq.~(\ref {eq:newOSR2}) determines
$\rho(t)$ in terms of $\rho (0),$ i.e., at the special time $t=0$ where the
system and the bath are in a product state. 

%
%
We now show that 
explicit use of a coarse-graining over time, together with the
above-mentioned assumptions, leads one naturally from the OSR 
evolution equation, Eq.~(\ref{eq:newOSR2}) to the SME. 
We note that it is of course possible to derive the SME with other methods,
(such as adding an infinite bath~\cite{Davies:74,Davies:76}) given the appropriate
assumptions.  Our goal here however is not so much to rederive the SME, as
to specifically establish a route from the non-Markovian OSR to the 
Markovian SME. 
 
At this point it is useful to introduce a time-scale $\tau $ for the bath
``memory'' (whose definition will be made more precise below) and to
coarse-grain the evolution of the system in terms of this time scale:

\begin{equation}
\rho _{j}=\rho (j\tau );\quad \chi _{\alpha \beta ;j}=\chi _{\alpha \beta
}(j\tau );\quad j\in {\bf N}.
\end{equation}
Further, rewriting the OSR Eq.~(\ref{eq:newOSR}) as $\rho (t)={\bf \Lambda }
(t)\rho (0)$ and defining $\tilde{{\tt L}}(t)$ through ${\bf \Lambda } (t)= 
{\rm T}\exp \left[ \int_{0}^{t}\tilde{{\tt L}}(s)ds\right] $ we find that

\begin{equation}
{\frac{\partial \rho (t)}{\partial t}}=\tilde{{\tt L}}(t)[\rho (t)].
\label{eq:OSRtrue}
\end{equation}
Defining $\tilde{{\tt L}}_{j}=\int_{j\tau }^{(j+1)\tau }\tilde{{\tt L}}(s)ds$
, with $\tau n=t$, we have

\begin{equation}
\int_{0}^{t}\tilde{{\tt L}}(s)ds=\tau \sum_{j=0}^{n-1}\tilde{{\tt L}}_{j}.
\end{equation}
Next we will make the assumption that on the time scale of the bath $\tau $,
the evolution generators $\tilde{{\tt L}}(t)$ commute in the ``average''
sense that $\left[ \tilde{{\tt L}}_{j},\tilde{{\tt L}}_{k}\right] =0,\forall
j,k$. Physically, we imagine this operation as arising from the
``resetting'' of the bath density operator over the time-scale $\tau $.
Under this assumption, the evolution of the system is Markovian when $t\gg
\tau $:

\begin{equation}
{\bf \Lambda}(t)=\prod_{j=0}^{n-1} \exp \left[ \tau \tilde{{\tt L}}_j \right]
.
\end{equation}
Further, under the discretization of the evolution, this product form of the
evolution implies that

\begin{equation}
\rho_{j+1}=\exp \left[ \tau \tilde{{\tt L}}_j \right] [\rho_j] .
\end{equation}
In the limit of $\tau \ll t$ we expand this exponential, to find that

\begin{equation}
{\frac{\rho_{j+1}-\rho_j }{\tau}}= \tilde{{\tt L}}_j [ \rho_j ].
\label{eq:rhodot}
\end{equation}
This equation is simply a discretization of Eq.~(\ref{eq:OSRtrue}) under the
assumption that $\tau \ll \theta$, where $\theta$ is the time-scale of
change for the system density matrix. Notice in particular that the RHS of
Eq.~(\ref{eq:rhodot}) contains the {\it average} value of $\tilde{{\tt L}}
(t) $ over the interval. Now, from the OSR evolution equation (\ref
{eq:newOSR2}), we know the explicit form of $\tilde{{\tt L } }(t)$ over the
first interval from $0$ to $\tau$. Discretizing over this interval we find
that

\begin{equation}
{\frac{\rho_{1}-\rho_0 }{\tau}} = - {\frac{i }{\hbar}} \left [ \left \langle 
\dot{{\bf S}} \right \rangle,\rho_0 \right ] + {\frac{1 }{2}}
\sum_{\alpha,\beta=1}^M \left \langle \dot{\chi}_{\alpha \beta} \right
\rangle \left( [{\bf K} _\alpha, \rho_0 {\bf K}_\beta^\dagger] + [{\bf K}
_\alpha \rho(0),{\bf K} _\beta^\dagger] \right) \equiv \tilde{{\tt L}}_0 [
\rho_0 ],
\end{equation}
where

\begin{equation}
\left\langle X\right\rangle \equiv {\frac{1}{\tau }}\int_{0}^{\tau }X(s)ds.
\label{eq:t-ave}
\end{equation}
Thus, in the sense of the coarse graining above we have arrived at an
explicit form for $\tilde{{\tt L}}_{0}$. Consider the evolution beyond this
first interval. Deriving an explicit form for $\tilde{{\tt L}}_{1}$ and for
higher terms is now impossible because Eq.~(\ref{eq:newOSR2}) gives the
evolution in terms of $\rho (0)$. However, since we have made the assumption
that the bath ``resets'' over the time-scale $\tau $, we expect the bath to
interact with the system in the same manner over every $\tau $-length
coarse-grained interval. This is equivalent to assuming that $\tilde{{\tt L}}
_{i}=\tilde{{\tt L}}_{0},\forall i$ (which of course is the most trivial way
of satisfying the Markovian evolution condition $[\tilde{{\tt L}}_{i},\tilde{
{\tt L}}_{j}]=0,\forall i,j$). Then, using Eq.~(\ref{eq:rhodot}), one is led
to the well known form of the semigroup equation of motion:

\begin{equation}
\frac{\partial \rho(t)}{\partial t}= - {\frac{i }{\hbar}} [ \left \langle{\ 
\dot{{\bf S}}} \right \rangle,\rho(t)] + {\frac{1 }{2}} \sum_{\alpha,
\beta=1}^M \left \langle {\dot{\chi}}_{\alpha \beta} \right \rangle \left( [ 
{\bf K}_\alpha, \rho(t) {\bf K}_\beta^\dagger] + [{\bf K}_\alpha \rho(t), 
{\bf K}_\beta^\dagger] \right)  \label{eq:newOSR3}
\end{equation}
(under the natural identification of the ${\bf K}$'s with the ${\bf F}$'s of
the SME).

We can write this equation of motion in an alternative form which
distinguishes between the system and bath contributions to the Liouvillian
evolution. Because Eq.~(\ref{eq:newOSR2}) is linear in the $\chi_{\alpha
\beta}(t)$ matrix, one can calculate $\chi _{\alpha \beta }^{(0)}(t)$ for
the isolated system and hence define the new terms which come about from the
coupling of the system to the bath:

\begin{equation}
\chi _{\alpha \beta }(t)=\chi _{\alpha \beta }^{(0)}(t)+\chi _{\alpha \beta
}^{(1)}(t).
\end{equation}
The terms which correspond to the isolated system will therefore produce a normal 
$-(i/\hbar )[{\bf H},\rho (t)]$ Liouville term in Eq.~(\ref{eq:newOSR3}).
Thus Eq.~(\ref{eq:newOSR3}) can be rewritten as

\begin{equation}
\frac{\partial \rho (t)}{\partial t}=-{\frac{i}{\hbar }}\left[ {\bf H}
+\left\langle {\dot{{\bf S}}}^{(1)}\right\rangle ,\rho (t)\right] +{\frac{1}{
2}}\sum_{\alpha ,\beta =1}^{M}\left\langle {\dot{\chi}}_{\alpha \beta
}^{(1)}\right\rangle \left( [{\bf K}_{\alpha },\rho (t){\bf K}_{\beta
}^{\dagger }]+[{\bf K}_{\alpha }\rho (t),{\bf K}_{\beta }^{\dagger }]\right)
,  \label{eq:newOSR4}
\end{equation}
which with the identification of $\left\langle {\dot{\chi}}_{\alpha \beta
}\right\rangle $ with $a_{\alpha \beta }$, and ${\bf K}_{\alpha }$ with $
{\bf F}_{\alpha }$, is equivalent to Eqs.({\ref{eq:SG1})-(\ref{eq:SG}),
except for the presence of the second term derived from $\left\langle {\dot{{
\ {\bf S}}}}^{(1)}\right\rangle $ in the Liouvillian. This second term
inducing unitary dynamics on the system, $\left\langle {\dot{{\bf S}}}
^{(1)}\right\rangle $, is referred to as the {\em Lamb shift}. It explicitly
describes the effect the bath has on the unitary part of the system dynamics
and ``renormalizes'' the system Hamiltonian. It is often implicitly assumed
to be present in Eq.~(\ref{eq:SG1}) \cite{Beck:93}. }

In summary, we have shown in this Section how coarse-graining the evolution
over the bath time-scale $\tau $ allows one to understand the connection
between the OSR and the semigroup evolution.  Specifically, we have
made the assumptions that (i) the time-scale for the evolution of the
system density matrix is much larger than the time-scale for the
resetting of the bath ($\tau \gg \theta$), (ii) the evolution of the
system should be Markovian ($[\tilde{{\tt L}}_{i},\tilde{
{\tt L}}_{j}]=0,\forall i,j$), and (iii) the bath resets to the same
state so that the system evolution is the same over every coarse
graining ($\tilde{{\tt L}}_i=\tilde{{\tt L}}_0,\forall i$).  This last 
assumption can be relaxed, and replaced by an ensemble average taken
over the different states to which the bath resets, i.e. 
$\tilde{{\tt L}}_i=\langle\langle \tilde{{\tt L}}_i \rangle\rangle$.  The importance of Eq.~(\ref
{eq:newOSR2}) lies in the fact that it allows one to pinpoint the exact
point at which the assumption of Markovian dynamics are made and further,
due to the general likeness of its form to the SME, provides an easily
translatable connection when going from the non-Markovian OSR to the
Markovian SME. Notice also that the assumption of Markovian dynamics
introduces an arrow of time in the evolution of the system through the
ordering of the environmental states: the system evolves through time in the
direction of each successive resetting of the bath.

\section{Short-Time Expansions of the Memory Fidelity in the OSR}

\label{short}

The {\em mixed-state memory fidelity} \cite{Jozsa:94}

\begin{equation}
F_{{\rm m}}(t)={\rm Tr}[\rho (0)\rho (t)]  \label{eq:Fmdef}
\end{equation}
is a good measure of the degree to which a system serves as a perfect
quantum memory. $F_{{\rm m}}(t)$ is the mixed state analog of the survival
probability for a pure state wavefunction. When the initial preparation is
pure, a perfect, noiseless quantum memory will have $F_{{\rm m}}(t)=1$, but
in the noisy case $0\leq F_{{\rm m}}(t)\leq 1$. If one starts out in a mixed
state then $F_{{\rm m}}(0)<1$, and it is usually necessary to resort to some
kind of purification \cite{Bennett:1}. We will consider here only short-time
expansions of the fidelity, since it is known that using QECC it is possible
to restore the coherence of a quantum system as long as corrections are
applied sufficiently frequently \cite{Knill:98}. Thus, we perform a power
expansion of the fidelity in time \cite{Duan:97a}

\begin{equation}
F_{{\rm m}}(t)=\sum_{n}\frac{1}{n!}\left( \frac{t}{\tau _{n}}\right) ^{n}
\end{equation}
where the {\em decoherence rates} are defined as

\begin{equation}
\frac{1}{\tau _{n}}=\left\{ {\rm Tr}[\rho (0)\rho ^{(n)}(0)]\right\} ^{1/n},
\end{equation}
and $\rho ^{(n)}$ denotes the $n^{{\rm th}}$ time derivative of the density
matrix.

\subsection{First Order Decoherence Rate in the OSR}

\label{tau1}

Throughout the literature on decoherence there abound many examples of
non-zero first order decoherence rates (e.g.\cite
{Zanardi:98a,Vitali:98,Zanardi:98}). Specific attention has been given to
maximizing this time scale in order to maintain long-lived coherent states.
We therefore pose the question here, how do the first order decoherence
rates for non-Markovian evolution behave within the OSR? The first order
decoherence rate is given by 
\begin{equation}
{\frac{1 }{\tau_1}}= {\rm Tr}[\rho(0)\dot{\rho}(0)].
\label{eq:tau1_inv}
\end{equation}
%
%
We note that by substituting in the reduced density matrix, Eq.~(\ref{eq:dyna}),
and evaluating the derivative at $t=0$, 
we are immediately led to the vanishing of the
first order decoherence rate from the cyclic property of the trace:
\begin{equation}
{\frac{1 }{\tau_1}}= {\rm Tr}_S[\rho(0) {\rm Tr}_B[-i{\bf H}_{SB}
\rho(0) +i \rho(0) {\bf H}_{SB}]] = 0.
\end{equation}
Therefore for the general non-Markovian dynamics, first order decoherence 
rates are rigorously zero, provided that ${\bf H}_{SB}$ is finite.
What is not obvious from this simple manipulation is how a
coarse-graining procedure can lead to the commonly encountered non-vanishing 
first order decoherence rates.  This therefore suggests employing
the ``pre-coarse-grained'' OSR, Eq.~(\ref{eq:newOSR2}), for the derivative,
and then carrying out the
specific coarse-graining procedure outlined in the previous section
on Eq.~(\ref{eq:tau1_inv}), in order to understand how non-vanishing first 
order decoherence rates can arise. 

Using Eq.~(\ref{eq:newOSR2}) the first order decoherence rate becomes

\begin{equation}
{\frac{1}{\tau _{1}}}={\rm Tr}\left[ \rho (0)\left( -{\frac{i}{\hbar }}[\dot{
{\bf S}}(0),\rho (0)]+{\frac{1}{2}}\sum_{\alpha ,\beta =1}^{M}\dot{\chi}
_{\alpha \beta }(0)\left( [{\bf K}_{\alpha },\rho (0){\bf K}_{\beta
}^{\dagger }]+[{\bf K}_{\alpha }\rho (0),{\bf K}_{\beta }^{\dagger }]\right)
\right) \right] .  \label{eq:1/tau1}
\end{equation}
Using the decomposition of the Kraus operators, Eq.~(\ref{eq:OSR}), and
knowing that ${\bf U}(0)={\bf I}_{S}\otimes {\bf I}_{B}$, we find that ${\bf 
A}_{i}(0)=\sqrt{\nu }{\bf I}_{S}\delta _{i,(\nu ,\nu )}$. Thus, since the $
{\bf K}_{\alpha }$'s form a linearly independent basis, it follows, using
Eq.~(\ref{eq:A-F}), that the expansion coefficients must be

\begin{equation}
b_{i\alpha}(0)= \delta_{\alpha 0} \sqrt{\nu} \delta_{i,(\nu,\nu)}.
\end{equation}
By direct evaluation,

\begin{equation}
\dot{\chi}_{\alpha \beta }(0)=\sum_{\nu }\sqrt{\nu }\left[ \left( \delta
_{\alpha 0}\dot{b}_{(\nu ,\nu ),\beta }^{\ast }(0)+\delta _{\beta 0}\dot{b}
_{(\nu ,\nu ),\alpha }(0)\right) \right] ,  \label{eq:chicond}
\end{equation}
which implies the vanishing [as long as $\dot{b}_{(\nu ,\nu ),\alpha }(0)$
remains finite] in Eq.~(\ref{eq:1/tau1}) of every term except ${\rm Tr}[\rho
(0)[\dot{{\bf S}}(0),\rho (0)]]$. However, this in turn vanishes by cyclic
permutation of the trace. Thus we see that within the OSR, {\em the first
order decoherence rate is always zero when the $\dot{b}_{(\nu ,\nu ),\alpha
}(0)$ remain finite}. To determine the significance of this restriction we
choose as a basis for the Kraus operators a set of $M$ ${\bf K}_{\alpha }$'s
which form a Lie algebra and hence have a suitably defined inner product
[Eq.~(\ref{eq:IP})]:

\begin{eqnarray}
{\rm Tr} \left[ {\bf K}_\alpha {\bf K}_\beta^\dagger\right] = \left \{ 
\begin{array}{c}
\delta_{\alpha\beta} \quad {\rm for} \quad \alpha \geq 1,\beta \geq 1 \\ 
N \delta_{\beta 0} \delta_{\alpha 0} \quad {\rm otherwise.}
\end{array}
\right.
\end{eqnarray}
We then find using the definition of the Kraus operators, Eq.~(\ref{eq:Amunu}
),

\begin{equation}
{b}_{(\nu \nu),\alpha}(t)={\rm Tr} \left[ {\bf K}_\alpha ^\dagger \sqrt{\nu}
\langle \nu | {\bf U}(t) | \nu \rangle \right].
\end{equation}
Differentiating this and recalling that ${\bf U}(t) = \exp(-i {\bf H}_{SB}
t/\hbar) $, we find that

\begin{equation}
\dot{b}_{(\nu \nu ),\alpha }(0)=-{\frac{i}{\hbar }}{\rm Tr}\left[ {\bf K}
_{\alpha }^{\dagger }\sqrt{\nu }\langle \nu |{\bf H}_{SB}(0)|\nu \rangle 
\right] .
\end{equation}
Thus in order for $\dot{b}_{(\nu \nu ),\alpha }(0)$ to remain finite, $
\langle \nu |{\bf H}_{SB}(0)|\nu \rangle $ must be finite. Hence our
conclusion that the first order decoherence rate vanishes in the OSR is
valid for any {\em finite} total Hamiltonian (by which we mean its matrix
elements), and conversely, any finite total Hamiltonian will have zero first
order decoherence rate \cite{comment}.

Examination of our derivation of the SME, Eqs.~(\ref{eq:newOSR3}) and (\ref
{eq:newOSR4}), now shows how non-zero first order decoherence rates can
arise when the evolution is considered to be Markovian. In the derivation of
the semigroup equation in the Markovian limit we made the assumption that
the matrices $\dot{\chi}_{\alpha \beta }(t)$ can be identified with the
constant matrices $a_{\alpha \beta }$ of the semigroup equation, Eq.~(\ref
{eq:SG}). However, when this is done, the matrix elements $\dot{\chi}
_{\alpha \beta }(0)$ in Eq.~(\ref{eq:1/tau1}) are replaced by their
time-averaged values, for which the relation (\ref{eq:chicond}) no longer
applies. Hence, in general, the first order decoherence rates are necessarily not zero
when the Markovian coarse-graining is applied. For a finite total
Hamiltonian ${\bf H}_{SB}$, non-zero first order rates are therefore seen to
be an artifact of the Markovian assumption, and their appearance emphasizes
the delicate nature of the transition to the Markovian regime.

\subsection{Example: Phase Damping}

\label{PD}

The restriction to a finite total Hamiltonian above may at first sight seem
obvious. However, consider, for instance, the often quoted example of phase
damping of a qubit. In this case, it would appear that there is a finite
first order decoherence rate. Yet, it is often presented within the OSR \cite
{Chuang:97,Knill:97b}, which, as we have just shown above, would predict 
{\em zero} first order decoherence rates for any non-singular
Hamiltonian.
In this example, the Kraus operators are given by {\cite{Chuang:97} }

\begin{eqnarray}
{\bf A}_0=\left( 
\begin{array}{cc}
1 & 0 \\ 
0 & e^{-\lambda t}
\end{array}
\right) \quad {\bf A}_1= \left( 
\begin{array}{cc}
0 & 0 \\ 
0 & \sqrt{1 - e^{-2\lambda t}}
\end{array}
\right) ,  \label{eq:KrausPh}
\end{eqnarray}
and a simple calculation using these operators yields a minimum first order
decoherence rate of $1/\tau_1=-\lambda/2$. How can this be?

To resolve this dichotomy, we consider how the above phase damping Kraus
operators are generated from the unitary dynamics of a qubit system $S$ and
a qubit bath $B$. The evolution operator

\begin{equation}
{\bf U}(t)=\left( 
\begin{array}{cccc}
1 & 0 & 0 & 0 \\ 
0 & e^{-\lambda t} & 0 & \sqrt{1 - e^{-2 \lambda t}} \\ 
\ 0 & 0 & 1 & 0 \\ 
0 & -\sqrt{1-e^{-2\lambda t}} & 0 & e^{-\lambda t}
\end{array}
\right) 
\begin{array}{c}
|\! \downarrow 0 \rangle \\ 
|\! \downarrow 1 \rangle \\ 
|\! \uparrow 0 \rangle \\ 
|\! \uparrow 1 \rangle
\end{array}
\end{equation}
[where the first qubit represents the bath ($|\! \uparrow \rangle, |\!
\downarrow \rangle$) and the second represents the system ($|0
\rangle,|1\rangle$) as denoted in the columns above] with the bath initially
in the state $|\! \downarrow \rangle$, immediately gives the Kraus operators
of Eq.~(\ref{eq:KrausPh}). Now, it is easy to calculate the Hamiltonian
which generates this evolution, [using ${\bf H}_{SB}(t)=i \hbar {\frac{d{\bf 
U}(t) }{dt}} {\bf U}^\dagger(t)$]:

\begin{equation}
{\bf H}_{SB}(t)= \left( 
\begin{array}{cccc}
0 & 0 & 0 & 0 \\ 
0 & 0 & 0 & -g(t) \\ 
0 & 0 & 0 & 0 \\ 
0 & g(t) & 0 & 0
\end{array}
\right),
\end{equation}
where 
\begin{equation}
g(t)=i\hbar {\frac{\gamma e^{-\gamma t} }{\sqrt{1 - e^{-2 \gamma t}}}}
\end{equation}
However, we see that as $t \rightarrow 0$, $g(t) \rightarrow \infty$. Thus,
in this simple example, we find that at $t=0$, the Hamiltonian becomes
singular. This illustrates our claim that first order decoherence rates in
the OSR are the result of an infinite Hamiltonian, and do not contradict the
general OSR result of zero rates for finite Hamiltonians.  

The diverging Hamiltonian in this example is in fact equivalent to {\em 
non-closedness} of the system $S+B$. It is well known that phase damping can
be generated by a model of random phase-kicks \cite{Silbey:89}, which
implies an {\em external} random force, i.e., that the system $S+B$ is in
fact not closed. Since this is in contradiction to our initial assumptions
(Sec.~\ref{OSR}), it should not come as a surprise that a non-vanishing
first order decoherence rate is found in this situation. A similar
divergence will result of course from a bath with an infinite number of
degrees of freedom \cite{Davies:74,Davies:76}.  The OSR phase damping example thus can still be
used (as is commonly done in the analysis of quantum error correction) under 
the caveat that one cannot claim that it arises from a finite closed
system (S+B).

\section{Effect of Symmetry Breaking Perturbations On Memory Fidelity of
Decoherence-Free Subspaces}

\label{SPB}

Our discussion in the previous sections was completely general, dealing with
the decoherence of the entire system Hilbert space. We now restrict our
attention to the behavior of the fidelity in DF subspaces. We first briefly
summarize the basic theory of DF subspaces and then generalize the first
order stability results obtained within the SME in Ref.~\cite{Lidar:PRL98}.

\subsection{Theory of Decoherence-Free Subspaces: Markovian {\it vs}
non-Markovian Approach}

\label{DFS}

Recently, conditions for the existence of decoherence free subspaces within
the framework of the Markovian SME approach \cite{Zanardi:98a,Lidar:PRL98}
and in a non-Markovian \cite{Zanardi:97a} setting were derived. We first
clarify here the connection between the SME and the non-Markovian results.

In the SME approach it was shown that a necessary and sufficient condition
for decoherence free dynamics (${\tt L}_{D}[\tilde{\rho}(t)]=0$) in a
subspace $\tilde{{\cal H}}={\rm Span}[\{|\tilde{\imath}\rangle
_{i=1}^{N_{0}}\}]$ is that all of the basis states $|\tilde{\imath}\rangle $
satisfy the condition

\begin{equation}
{\bf F}_{\alpha }|\tilde{\imath}\rangle =c_{\alpha }|\tilde{\imath}\rangle
\quad \forall \alpha ,\tilde{\imath}  \label{eq:SGcond}
\end{equation}
where the ${\bf F}_{\alpha }$'s are the error generators in the semigroup
Eq.~(\ref{eq:SG}). Since the $\{{\bf F}_{\alpha }\}$ form a Lie algebra $
{\cal L}$, this condition has a simple group-theoretic interpretation,
namely, the DF states are the {\em singlets} of ${\cal L}$, i.e., they are
the states that transform according to the 1-dimensional representations of $
{\cal L}$ . From Sec.~\ref{OSR-SG}, and in particular Eq.~(\ref{eq:newOSR3}), it follows that these error generators become identical to the $\{{\bf K}
_{\alpha }\}$ (the basis operators in the fixed expansion of the Kraus
operators) when the short-time averaging approximations leading from the OSR
to the SME are made.

Within the framework of non-Markovian evolution, it has likewise been shown 
\cite{Zanardi:97a,Lidar:98QECC-DFS} that a necessary and sufficient
condition for decoherence free dynamics over a similar subspace $\tilde{
{\cal H}}={\rm Span}[\{|\tilde{\imath}\rangle _{i=1}^{N_{0}}\}]$ is that all
of the basis states $|\tilde{\imath}\rangle $ satisfy the condition

\begin{equation}
{\bf S}_{\alpha }|\tilde{\imath}\rangle =c_{\alpha }|\tilde{\imath}\rangle
\quad \forall \alpha ,\tilde{\imath}  \label{eq:DFOSR}
\end{equation}
where ${\bf S}_{\alpha }$'s (system operators) are defined by the
interaction Hamiltonian

\begin{equation}
{\bf H}_{I}=\sum_{\alpha }{\bf S}_{\alpha }\otimes {\bf B}_{\alpha }.
\label{eq:splitHI}
\end{equation}
The ${\bf B}_{\alpha }$ are bath operators. The fixed-basis OSR
equation (\ref{eq:newOSR}) sheds light on the relationship between
these two DF
conditions. To the extent that the error generators ${\bf F}_{\alpha }$ can
be derived from expansion of the Kraus operators with a subsequent
short-time averaging approximation, the DF condition given for the
non-Markovian dynamics are more general than that given by the semigroup
approach. We notice that in the limit of small averaging time $\tau $, the
error generators ${\bf F}_{\alpha }$'s will, in fact, correspond directly to
the ${\bf S}_{\alpha }$'s. This can be seen by expanding the full evolution
operator ${\bf U}(t)$ to first order in $\tau $:

\begin{equation}
{\bf U}(\tau )={\bf I}-{\frac{i}{\hbar }}{\bf H}_{SB}\tau +O(\tau ^{2}).
\end{equation}
To this order, the Kraus operators ${\bf A}_{i}(t)$ [see Eq.~(\ref{eq:Amunu}
)] will only contain terms which correspond to terms that appear in the
Hamiltonian ${\bf H}_{SB}$. These are ${\bf H}$, the system Hamiltonian; $
{\bf I}$, the identity; and the ${\bf S_{\alpha }}^{\prime }$s from Eq.~(\ref
{eq:splitHI}). Terms corresponding to ${\bf H}$ will result in unitary
evolution of the system, while ${\bf I}$ is removed from the set of Kraus
operators (since these are considered error generators) via our derivation
of the SME. To first order in time, therefore the only error generators are
the ${\bf S}_{\alpha }$'s. Thus in the case of small averaging time $\tau$
we see that the two DF conditions are exactly equivalent.

However, it is important to note that the SME approach has other advantages.
Thus in many cases it is either impractical or undesirable to derive the $\{ 
{\bf F} _{\alpha }\}$ from a short-time expansion of the type discussed in
Sec.~\ref{OSR-SG}. In fact, in Lindblad's axiomatic approach \cite
{Lindblad:76}, the $\{{\bf F}_{\alpha }\}$ are the primary objects and they
do not follow from an expansion of a unitary operator. While the $\{ {\bf F}
_{\alpha }\}$ are often identified heuristically from a factorization such
as in Eq.~(\ref{eq:splitHI}) \cite{Kosloff:97}, in some cases (notably
strong coupling) one simply cannot clearly separate system and bath in the
form assumed in that equation. In this sense, then, the SME provides greater
generality than the non-Markovian approach within the Hamiltonian
representation subject to Eq.~(\ref{eq:splitHI}). Motivated by this aspect,
we take condition (\ref{eq:SGcond}) to be necessary and sufficient for DF
subspaces.

Further, one should note that while the semigroup DF condition, Eq.~(\ref
{eq:SGcond}), guarantees that the evolution of the system will be unitary,
the system may still be subject to unitary evolution induced by the bath in
the form of the Lamb shift. Such bath-induced evolution, although it does
not introduce decoherence, is undesirable in the course of a quantum
computation. To the extent that we desire the DF subspace to serve as the
basis for a quantum computer, we therefore must impose one of two conditions
on the DF subspace. These conditions are: (1) suitable control over the
system is obtained so that the Lamb shift term can be canceled out, or, (2)
the Lamb shift does not induce dynamics on the subspace, i.e., ${\bf H}_{
{\rm Lamb}}|\tilde{\imath}\rangle =h_{l}|\tilde{\imath}\rangle $. Under the
first condition, the DF subspace is not reduced in dimension. However this
condition may be physically impossible to realize. The second condition does
not make any assumptions about the amount of control one has over the system
Hamiltonian, but it may have the undesirable effect of causing a reduction
in the size of the DF subspace. Similar conclusions regarding the effect of
the Lamb shift hold for the non-Markovian case.

Finally, in both the SME representation and the OSR, the implicit assumption
has been made that the system Hamiltonian does not induce the evolution of
states from within DF subspace to states outside of the DF subspace: 
\begin{equation}
{\bf H}|\tilde{\imath}\rangle =\sum_{j}h_{ij}|\tilde{j}\rangle .
\end{equation}
It is a simple matter to satisfy this additional condition in the SME.
However, this is not the case in the OSR due to the absence of an explicit
appearance of ${\bf H}$.

We now proceed to the question of stability of DF subspaces in the
non-Markovian and Markovian cases. The calculations in the next two
subsections are rather tedious, and the reader who is not interested in the
details may wish to skip directly to the results for the memory fidelity,
summarized in Table I.

\subsection{Stability of the Memory Fidelity in the Non-Markovian Case}

\label{SBP-OSR}

Consider the addition to a DF subspace of new perturbing terms in the
interaction Hamiltonian: ${\bf H}_{SB}^{\prime }={\bf H}_{SB}+\epsilon {\bf 
H }_{I}^{\prime }$. In this case, we find that to first order in $\epsilon $
, the new full evolution operator is given by

\begin{eqnarray}
{\bf U}^{\prime }(t) &=& \sum_{n=0}^\infty \frac{(-it/\hbar )^{n}}{n!}
\left( {\bf H}_{SB} + \epsilon {\bf H}_{I}^{\prime } \right)^n  \nonumber \\
&=& {\bf U}(t) + \sum_{k=1}^\infty \epsilon^k \sum_{n=k}^{\infty }\frac{
(-it/\hbar )^{n}}{n!} f^{(k)}_{n}({\bf H}_{SB},{\bf H}_{I}^{\prime }),
\end{eqnarray}
where

\begin{eqnarray}
f_{1}^{(1)}({\bf H}_{SB},{\bf H}_{I}^{\prime }) &=&{\bf H}_{I}^{\prime } 
\nonumber \\
f_{2}^{(1)}({\bf H}_{SB},{\bf H}_{I}^{\prime }) &=&{\bf H}_{SB}{\bf H}
_{I}^{\prime }+{\bf H}_{I}^{\prime }{\bf H}_{SB}  \nonumber \\
f_{3}^{(1)}({\bf H}_{SB},{\bf H}_{I}^{\prime }) &=&{\bf H}_{SB}^{2}{\bf H}
_{I}^{\prime }+{\bf H}_{SB}{\bf H}_{I}^{\prime }{\bf H}_{SB}+{\bf H}
_{I}^{\prime }{\bf H}_{SB}^{2}  \nonumber \\
f_{2}^{(2)}({\bf H}_{SB},{\bf H}_{I}^{\prime }) &=&{{\bf H}_{I}^{\prime }}
^{2}  \nonumber \\
f_{3}^{(2)}({\bf H}_{SB},{\bf H}_{I}^{\prime }) &=&{\bf H}_{SB}{{\bf H}
_{I}^{\prime }}^{2}+{\bf H}_{I}^{\prime }{\bf H}_{SB}{\bf H}_{I}^{\prime }+{
\ \ {\bf H}_{I}^{\prime }}^{2}{\bf H}_{SB},
\end{eqnarray}
etc. Here ${\bf U}(t)$ is the unperturbed evolution operator. From Eqs.~(\ref
{eq:Amunu}) and (\ref{eq:A-F}) we thus see that to first order in $\epsilon $
the operators ${\bf K}_{\alpha }$ in which the Kraus operators are expanded,
will have new terms due to $\{f_{n}^{(1)}\}$, hereby denoted by $\left\{
\epsilon {\bf G}_{p}\right\} _{p=1}^{P}$, which are proportional to $
\epsilon $. These terms modify the evolution over the DF subspace [Eq.~(\ref
{eq:newOSR})] so that $\partial \tilde{\rho}/\partial t\mapsto \partial 
\tilde{\rho}^{\prime }/\partial t=\partial \tilde{\rho}/\partial t+{\tt L}
^{\prime }(t)[\tilde{\rho}(0)]$, where:

\begin{eqnarray}
{\tt L}^{\prime }(t)[\tilde{\rho}(0)]\equiv &-& {\frac{i}{\hbar }}[{\bf S}
^{\prime }(t),\tilde{\rho}(0)]+{\frac{1}{2}}\sum_{\alpha
=1}^{M}\sum_{p=1}^{P}\chi _{\alpha p}(t){\tt L}_{{\bf K}_{\alpha },\epsilon 
{\bf G}_{p}}[\tilde{\rho}(0)]+\chi _{\alpha p}^{\ast }(t){\tt L}_{\epsilon 
{\bf G}_{p},{\bf K}_{\alpha }}[\tilde{\rho}(0)]  \nonumber \\
&+& O(\epsilon ^{2}).  \label{eq:newterms}
\end{eqnarray}
Here:

\begin{equation}
{\bf S}^{\prime}(t) = \epsilon {\frac{i \hbar}{2}} \sum_{p=1}^P \left[
\chi_{p 0}(t) {\bf G}_p - \chi_{0 p}(t) {\bf G}_p^\dagger \right] ,
\label{eq:defJp}
\end{equation}
and:

\begin{equation}
{\tt L}_{{\bf x},{\bf y}}[\rho ]\equiv \lbrack {\bf x},\rho {\bf y}^{\dagger
}]+[{\bf x}\rho ,{\bf y}^{\dagger }].  \label{eq:defL}
\end{equation}
Terms of $O(\epsilon ^{2})$, not written out explicitly in Eq.~(\ref
{eq:newterms}), include ${\tt L}_{\epsilon {\bf G}_{p},\epsilon {\bf G}_{q}}[
\tilde{\rho}(0)]$, ${\tt L}_{\epsilon ^{2}{\bf G}_{p},{\bf K}_{\alpha }}[
\tilde{\rho}(0)]$, etc. Assuming that $\epsilon \ll 1$, we may neglect these
terms. Now, for the purposes of argument, we will assume that the system has
perfect quantum {\em memory }over the DF subspace in the absence of the
perturbing error generators, i.e., $F_{{\rm m}}^{({\rm DF})}(t)={\rm Tr}[
\tilde{\rho}(0)\tilde{\rho}(t)]=1$. The perturbation, however, decreases the
fidelity below this perfect value. The modified memory fidelity can be
written formally as:

\[
F_{{\rm m}}^{\prime }(t)=1-\sum_{k=1}\sum_{n=k}\frac{1}{n!}\frac{t^{n}}{
\left( \tau _{n}^{(k)}\right) ^{n}}. 
\]
where $\tau _{n}^{(k)}$ represents the $O(\epsilon ^{k})$ contribution to
the $n^{{\rm th}}$ order decoherence time $\tau _{n}$. It was shown in Ref.~ 
\cite{Lidar:PRL98} that, within the SME, the term which is first order in
both the perturbation and time vanishes: $1/\tau _{1}^{(1)}=0$. This left
open the possibility of terms of order $O(\epsilon t^{2})$ and higher
spoiling the fidelity. Here we will generalize this result in the
non-Markovian case [the Markovian case will be dealt with in Sec.~(\ref
{SBP-SME})] and show that in fact $1/\tau _{n}^{(1)}=0$ for all $n$, so that
the entire $O(\epsilon )$ contribution vanishes, and only terms of order $
O(\epsilon ^{2}t^{2})$ can spoil the memory fidelity. For simplicity of
notation, since we are considering here only the $1/\tau _{n}^{(1)}$
decoherence rates, we drop the $(k)$ superscript from now on.

The {\em perturbed} decoherence rates are thus given by:

\begin{equation}
\left( \frac{1}{\tau _{n}}\right) ^{n}={\rm Tr}[\tilde{\rho}(0)\left\{ 
\tilde{\rho}^{\prime }(t)\right\} ^{(n)}]={\rm Tr}[\tilde{\rho}(0)\left\{ 
{\tt L}^{\prime }(t)[\tilde{\rho}(0)]\right\} _{t=0}^{(n-1)}].
\end{equation}
Using Eq.~(\ref{eq:newterms}) and noting that the terms involving ${\bf S}
^{\prime }$ vanish directly by permutation under the trace, we obtain to
first order in $\epsilon $:

\begin{equation}
\left( {\frac{1}{\tau _{n}}}\right) ^{n}={\frac{1}{2}}\left[ \sum_{\alpha
=1}^{M}\sum_{p=1}^{P}\chi _{\alpha p}^{(n-1)}(t){\rm Tr}[\tilde{\rho}(0){\tt 
L}_{{\bf K}_{\alpha },\epsilon {\bf G}_{p}}[\tilde{\rho}(0)]]+\chi _{\alpha
p}^{\ast (n-1)}(t){\rm Tr}[\tilde{\rho}(0){\tt L}_{\epsilon {\bf G}_{p},{\bf 
K}_{\alpha }}[\tilde{\rho}(0)]]\right] .  \label{eq:dts}
\end{equation}
To evaluate this, we need to know ${\bf K}_{\alpha }\tilde{\rho (0)}$. Now,
when we expand the Kraus operators about a fixed basis ${\bf K}_{\alpha }$
as in Eq.~(\ref{eq:A-F}), this basis will consist of all possible products
of the three terms ${\bf H}$, ${\bf S}_{\alpha }$, and ${\bf I}$ [recall the
definition of the Kraus operators, Eq.~(\ref{eq:Amunu})]. Assuming a perfect
quantum memory 
\begin{equation}
\lbrack {\bf H},\tilde{\rho}(0)]=0,  \label{eq:Hrho}
\end{equation}
we can commute ${\bf H}$ with $\tilde{\rho}(0)$ and, using the DF condition $
{\bf S}_{\alpha }|\tilde{\imath}\rangle =c_{\alpha }|\tilde{\imath}\rangle $
, for a given product of ${\bf S}_{\alpha }$'s and ${\bf H}$'s we can
replace each ${\bf S}_{\alpha }$ with its eigenvalue $c_{\alpha }$. Thus,
for example,

\begin{equation}
{\bf H}^{2}{\bf S}_{1}{\bf H}\tilde{\rho}={\bf H}^{2}{\bf S}_{1}\tilde{\rho} 
{\bf H}={\bf H}^{2}c_{1}\tilde{\rho}{\bf H}=c_{1}{\bf H}^{3}\tilde{\rho}.
\end{equation}
It follows that for a DF subspace in the OSR, the basis operators ${\bf K}
_{\alpha }$ will satisfy the condition 
\begin{equation}
{\bf K}_{\alpha }\tilde{\rho}(t)=d_{\alpha }{\bf H}^{m_{\alpha }}\tilde{\rho}
(t),
\end{equation}
with $m_{\alpha }$ an integer and $d_{\alpha }$ a real number. Using this
result, we then cycle the trace in Eq.~(\ref{eq:dts}), and again using $[ 
{\bf H},\tilde{\rho}(0)]=0$, we find for the first trace

\begin{eqnarray}
&&{\rm Tr}[\tilde{\rho}(0){\tt L}_{{\bf K}_{\alpha },\epsilon {\bf G}_{p}}[ 
\tilde{\rho}(0)]]  \nonumber \\
&=&\epsilon d_{\alpha }{\rm Tr}[2\tilde{\rho}(0){\bf H}^{m_{\alpha }}\tilde{
\rho}(0){\bf G}_{p}^{\dagger }-\tilde{\rho}(0){\bf G}_{p}^{\dagger }{\bf H}
^{m_{\alpha }}\tilde{\rho}(0)-\tilde{\rho}(0){\bf G}_{p}^{\dagger }\tilde{
\rho}(0){\bf H}^{m_{\alpha }}]  \nonumber \\
&=&0.
\end{eqnarray}
The second trace vanishes similarly. Thus we see that all orders of
decoherence rates must vanish to first order in $\epsilon $, i.e., $1/\tau
_{n}=0$. Examining Eq.~(\ref{eq:newterms}), it is clear that the second
order $\epsilon ^{2}$ term does not lead to a similar vanishing of the
traces. Therefore we have proved that: {\em under non-Markovian evolution,
DF subspaces are completely stable to first order under a symmetry breaking
perturbation}, where by ``completely'' we mean explicitly stable to all
orders of time.

\subsection{Stability of the Memory Fidelity in the Markovian Case}

\label{SBP-SME}

The stability of DF subspaces with respect to symmetry breaking
perturbations in the non-Markovian case derived above is an significant
extension of the stability derived in Ref.~\cite{Lidar:PRL98} for the
Markovian SME. However, the result presented in Ref.~\cite{Lidar:PRL98} only
examined the effect of a symmetry breaking perturbation on the first order
decoherence rate ($1/\tau _{1}^{(1)}=0$). Here we show that the stronger
non-Markovian result derived above ($1/\tau _{n}^{(1)}=0$ $\forall
n\geq 1$) also holds in the Markovian SME.

The effect of perturbing a DF subspace in the SME is the addition of new
error generators, hereby denoted by $\left\{ \epsilon {\bf G}_{p}\right\}
_{p=1}^{P}$, to the master equation (which was partially treated in Ref.~ 
\cite{Lidar:PRL98}) as well as a perturbing Lamb shift term in the master
equation (which was not treated in Ref.~\cite{Lidar:PRL98}). These terms
modify the SME [Eqs.~(\ref{eq:SG1})-(\ref{eq:SG})] so that ${\partial \rho
(t)/}\partial t={\tt L}^{\prime }[\rho (t)]$, with 
\begin{eqnarray}
{\tt L}^{\prime }[\rho (t)] &\equiv &{\tt L}[\rho (t)]-{\frac{i}{\hbar }}
[\epsilon {\bf H}_{{\rm Lamb}}^{\prime },\rho (t)]  \nonumber \\
&+&{\frac{1}{2}}\sum_{p=1}^{P}\sum_{\alpha =1}^{M}(g_{p\alpha }{\tt L}
_{\epsilon {\bf G}_{p},{\bf F}_{\alpha }}[\rho (t)]+g_{p\alpha }^{\ast }{\tt 
L}_{{\bf F}_{\alpha },\epsilon {\bf G}_{p}}[\rho (t)])+O(\epsilon ^{2}),
\label{eq:SMEnewterms}
\end{eqnarray}
where ${\tt L}[\rho (t)]$ is the unperturbed SME term given by Eq.~(\ref
{eq:SG1}), $\epsilon {\bf H}_{{\rm Lamb}}^{\prime }$ is the perturbing Lamb
shift, and ${\tt L}_{{\bf x},{\bf y}}[\rho (t)]$ is given by Eq.~(\ref
{eq:defL}). The perturbed decoherence rates are given by 
\begin{equation}
\left( \frac{1}{\tau _{n}}\right) ^{n}={\rm Tr}[\tilde{\rho}(0)\tilde{\rho}
^{(n)}(0)]={\rm Tr}[\tilde{\rho}(0)\left\{ ({\tt L}^{\prime })^{n}[\tilde{
\rho}(0)]\right\} ],
\end{equation}
where $({\tt L}^{\prime })^{n}[\tilde{\rho}(0)]={\tt L}^{\prime }[{\tt L}
^{\prime }[...{\tt L}^{\prime }[\tilde{\rho}(0)]]]$, $n$ times. To evaluate
this expression, recall (i) the DF condition ${\tt L}[\tilde{\rho}(0)]=0$
and (ii) that we are working only to first order in $\epsilon $. Now, for
simplicity, consider first $({\tt L}^{\prime })^{2}[\tilde{\rho}(0)]$, and
denote the second and third terms on the RHS of Eq.~(\ref{eq:SMEnewterms})
by {\tt A}$_{1}[\rho (t)]$ and {\tt A}$_{2}[\rho (t)]$. By the DF condition, 
${\tt L}[{\tt L}[\tilde{\rho}(0)]]=${\tt A}$_{1}[{\tt L}[\tilde{\rho}(0)]]=$ 
{\tt A}$_{2}[{\tt L}[\tilde{\rho}(0)]]=0$. Also, {\tt A}$_{i}[{\tt A}_{j}[ 
\tilde{\rho}(0)]]$ is of $O(\epsilon ^{2})$. Only the two terms with ${\tt L}
$ acting on {\tt A}$_{1}[\tilde{\rho}(0)]$ and {\tt A}$_{2}[\tilde{\rho}(0)]$
do not vanish. This reasoning generalizes easily for $n>2$, so we find that
to first order in $\epsilon $, 
\begin{eqnarray}
&&\left( \frac{1}{\tau _{n}^{(1)}}\right) ^{n}=  \nonumber \\
&&{\rm Tr}\left[ \tilde{\rho}(0){\tt L}^{n-1}\left[ -{\frac{i}{\hbar }}
[\epsilon {\bf H}_{{\rm Lamb}}^{\prime },\tilde{\rho}(0)]+{\frac{1}{2}}
\sum_{p=1}^{P}\sum_{\alpha =1}^{M}(g_{p\alpha }{\tt L}_{\epsilon {\bf G}
_{p}, {\bf F}_{\alpha }}[\tilde{\rho}(0)]+g_{p\alpha }^{\ast }{\tt L}_{{\bf F
} _{\alpha },\epsilon {\bf G}_{p}}[\tilde{\rho}(0)])\right] \right] .
\label{eq:SGnewdr}
\end{eqnarray}
Now, the DF condition, Eq.~(\ref{eq:SGcond}), implies that the DF error
generators commute with the DF density matrix: $[{\bf F}_{\alpha },\tilde{
\rho}(0)]=0$. We also again assume perfect quantum memory [Eq.~(\ref{eq:Hrho}
)]. Thus, for an arbitrary operator ${\bf A\in }{\cal A}({\cal H})$, we find
that we can commute the initial density matrix through the operator ${\tt L}$
: 
\begin{eqnarray}
{\tt L}[\tilde{\rho}(0){\bf A}] &=& -\frac{i}{\hbar }[{\bf H},\tilde{\rho}
(0) {\bf A}]+\frac{1}{2}\sum_{\alpha ,\beta =1}^{M}a_{\alpha \beta }(2{\bf F}
_{\alpha }\tilde{\rho}(0){\bf AF}_{\beta }^{\dagger }-{\bf F}_{\beta
}^{\dagger }{\bf F}_{\alpha }\tilde{\rho}(0){\bf A-}\tilde{\rho}(0){\bf AF}
_{\beta }^{\dagger }{\bf F}_{\alpha })  \nonumber \\
&=& -\frac{i}{\hbar }\tilde{\rho}(0)[{\bf H},{\bf A}]+\frac{1}{2}
\sum_{\alpha ,\beta =1}^{M}a_{\alpha \beta }\tilde{\rho}(0)(2{\bf F}_{\alpha
}{\bf AF} _{\beta }^{\dagger }-{\bf F}_{\beta }^{\dagger }{\bf F}_{\alpha }
{\bf A-AF} _{\beta }^{\dagger }{\bf F}_{\alpha })  \nonumber \\
&=& \tilde{\rho}(0){\tt L}[{\bf A}],  \nonumber
\end{eqnarray}
so that: 
\begin{equation}
{\tt L}^{(n-1)}[\tilde{\rho}(0){\bf A}]=\tilde{\rho}(0){\tt L}^{(n-1)}[{\bf 
A }];\qquad {\tt L}^{(n-1)}[{\bf A}\tilde{\rho}(0)]={\tt L}^{(n-1)}[{\bf A}] 
\tilde{\rho}(0).
\end{equation}
Examining the contribution to the decoherence rates from the Lamb shift
term, we thus find that 
\begin{equation}
\left( \frac{1}{\tau _{n}^{(1)}}\right) _{{\rm Lamb}}^{n}=-{\frac{i}{\hbar }}
{\rm Tr}[\tilde{\rho}(0)[{\tt L}^{(n-1)}[\epsilon {\bf H}_{{\rm Lamb}}], 
\tilde{\rho}(0)]]=0
\end{equation}
which vanishes by cycling under the trace. Next, the contribution to the
decoherence rates due to the symmetry breaking perturbing error generators
in Eq.~(\ref{eq:SGnewdr}) are given by 
\begin{equation}
\left( \frac{1}{\tau _{n}^{(1)}}\right) _{{\rm SBP}}^{n}={\frac{1}{2}}
\sum_{p=1}^{P}\sum_{\alpha =1}^{M}{\rm Tr}\left\{ \tilde{\rho}(0){\tt L}
^{(n-1)}[g_{p\alpha }{\tt L}_{\epsilon {\bf G}_{p},{\bf F}_{\alpha }}[\tilde{
\rho}(0)]+g_{p\alpha }^{\ast }{\tt L}_{{\bf F}_{\alpha },\epsilon {\bf G}
_{p}}[\tilde{\rho}(0)]]\right\} .  \label{eq:SGnewdrSBP}
\end{equation}
Expanding the first of these terms, 
\begin{eqnarray}
&&{\frac{\epsilon }{2}}\sum_{p=1}^{P}\sum_{\alpha =1}^{M}g_{p\alpha }{\rm Tr}
\left[ \tilde{\rho}(0){\tt L}^{(n-1)}[2{\bf G}_{p}\tilde{\rho}(0){\bf F}
_{\alpha }^{\dagger }-{\bf F}_{\alpha }^{\dagger }{\bf G}_{p}\tilde{\rho}
(0)- \tilde{\rho}(0){\bf F}_{\alpha }^{\dagger }{\bf G}_{p}]\right] 
\nonumber \\
&=&{\frac{\epsilon }{2}}\sum_{p=1}^{P}\sum_{\alpha =1}^{M}g_{p\alpha }{\rm 
Tr }\left[ \tilde{\rho}(0)\left( {\tt L}^{(n-1)}[2{\bf G}_{p}{\bf F}_{\alpha
}^{\dagger }]\tilde{\rho}(0)-{\tt L}^{(n-1)}[{\bf F}_{\alpha }^{\dagger } 
{\bf G}_{p}]\tilde{\rho}(0)-\tilde{\rho}(0){\tt L}^{(n-1)}[{\bf F}_{\alpha
}^{\dagger }{\bf G}_{p}]\right) \right]  \nonumber \\
&=&\epsilon \sum_{p=1}^{P}\sum_{\alpha =1}^{M}g_{p\alpha }{\rm Tr}\left[ 
\tilde{\rho}(0){\tt L}^{(n-1)}[[{\bf G}_{p},{\bf F}_{\alpha }^{\dagger }]] 
\tilde{\rho}(0)\right] \\
&=&\epsilon \sum_{p=1}^{P}\sum_{\alpha =1}^{M}g_{p\alpha }{\rm Tr}\left[ 
{\tt L}^{(n-1)}[\tilde{\rho}(0)[{\bf G}_{p},{\bf F}_{\alpha }^{\dagger }] 
\tilde{\rho}(0)]\right]
\end{eqnarray}
Using ${\bf F}_{\alpha }^{\dagger }\tilde{\rho}(0)=c_{\alpha }^{\ast }\tilde{
\rho}(0)$, we see that $\tilde{\rho}(0)[{\bf G}_{p},{\bf F}_{\alpha
}^{\dagger }]\tilde{\rho}(0)=\tilde{\rho}(0)({\bf G}_{p}c_{\alpha }^{\ast
}-c_{\alpha }^{\ast }{\bf G}_{p})\tilde{\rho}(0)$ and thus this term vanish.
Similar reasoning implies the vanishing of the second term in Eq.~(\ref
{eq:SGnewdrSBP}). Thus we have proven that $1/\tau _{n}^{(1)}=0$: {\em under
Markovian evolution, DF subspaces are completely stable to first order under
a symmetry breaking perturbation}\cite{zurek}. 

\section{The Dynamical Fidelity}

\label{dyna-F}

The results derived in the previous Section imply that DF subspaces are
robust to small perturbations when the DF subspace is operating as a quantum 
{\em memory}. In order to address what happens when perturbations are made
on the system as it evolves according to some desired quantum {\em 
computation}, we have to first define an analog of the mixed-state memory
fidelity for an evolving system. This is

\begin{equation}
F_{{\rm d}}(t)={\rm Tr}[\rho_U(t) \rho(t)] ,
\end{equation}
where $\rho_U(t)$ is the desired unitary evolution

\begin{equation}
\rho _{U}(t)={\bf U}_{S}(t)\rho (0){\bf U}_{S}^{\dagger }(t),\quad {\rm with}
\quad {\bf U}_{S}(t)=\exp \left[ -{\frac{i}{\hbar }}{\bf H}_{S}t\right] .
\end{equation}
Here ${\bf H}_{S}$ is the system Hamiltonian, and may include a ``program''
Hamiltonian which implements a quantum algorithm on the system. This {\em 
dynamical} fidelity is a good measure of the difference between the desired
evolution of the system and the actual, ``noisy'' evolution. Thus, $0\leq F_{
{\rm d}}(t)\leq 1$, with $F_{{\rm d}}(t)=1$ if and only if the evolution is
perfect, i.e., $\rho (t)=\rho _{U}(t)$.

The decoherence rates for the dynamical fidelity are defined in the same
manner as for the memory fidelity:

\begin{eqnarray}
F_{{\rm d}}(t)=\sum_{n}\frac{1}{n!}\left( \frac{t}{\bar{\tau}_{n}}\right)
^{n}:\quad \frac{1}{\bar{\tau}_{n}}=\left\{ {\rm Tr}[\left\{ \rho
_{U}(t)\rho (t)\right\} ^{(n)}]\right\} ^{1/n}.
\end{eqnarray}

\subsection{Markovian Case}

First we consider the dynamical fidelity within the context of the Markovian
limit, using the SME approach. We restrict our attention as before to DF
subspaces, so the density matrix $\tilde{\rho}$ satisfies Eq.~(\ref{eq:SG1})
with ${\tt L}_{D}[\tilde{\rho}(t)]=0$. We then imagine this DF subspace to
be perturbed by a symmetry-breaking perturbation: ${\tt L}_{D}\mapsto {\tt L}
_{D}^{\prime }$, where the perturbed density matrix satisfies the following
SME:

\[
\frac{\partial \tilde{\rho}}{\partial t}=-{\frac{i}{\hbar }}[{\bf H}_{S},
\tilde{\rho}(t)]+{\tt L}_{D}^{\prime }[\tilde{\rho}(t)].
\]
Similarly to Eq.~(\ref{eq:newterms}) (see also Ref.~\cite{Lidar:PRL98}), the
new terms in this SME are given by:

\begin{equation}
{\tt L}_D^{\prime}[\tilde{\rho}(t)]=\sum_{\alpha=1}^M \sum_{p=1}^P
\left(a_{\alpha p}{\tt L}_{{\bf F}_\alpha,\epsilon{\bf G}_p}[\tilde{\rho}
(t)] + a_{\alpha p}^* {\tt L}_{\epsilon{\bf G}_p,{\bf F}_\alpha}[\tilde{\rho}
(t)]\right) + O(\epsilon^2) .  \label{eq:SGnewstuff}
\end{equation}
The {\em perturbed} first order dynamical fidelity decoherence rate is given
by:

\begin{equation}
{\frac{1}{\bar{\tau}_{1}}}={\rm Tr}\left[ \tilde{\rho}_{U}(0)\frac{\partial 
\tilde{\rho}(0)}{\partial t}+\frac{\partial \tilde{\rho}_{U}(0)}{\partial t}
\tilde{\rho}(0)\right] .
\end{equation}
The first of these terms vanishes via the arguments given for the memory
fidelity (essentially, since ${\bf F}_{\alpha }\tilde{\rho}=c_{\alpha }
\tilde{\rho}$ by the DF subspace property). The second term also vanishes,
by permutation of the trace after using $\partial \rho _{U}(0)/\partial
t=-i\hbar \lbrack {\bf H}_{S},\rho (0)]$. Thus we find that {\em DF
subspaces are stable to first order in time also when the system is allowed
to evolve}.

Further, it is easy to see that the higher order dynamic fidelities now do
not vanish. For example, the second order dynamic decoherence rate contains
terms like $[{\bf H}_{S},\rho (0)]{\tt L}_{D}^{\prime }[\rho (t)]_{t=0}$,
which do not allow the simple permutation of the trace.

\subsection{Non-Markovian Case}

Is there an analogous result for the non-Markovian situation? We can address
this within the OSR. The arguments of Sec.~\ref{tau1} showed that the first
order decoherence rate will always vanish, and so according to the arguments
given above, this applies also to the dynamical fidelity. Examining the
second order decoherence rates, we find:

\begin{equation}
\left( {\frac{1}{\bar{\tau}_{2}}}\right) ^{2}={\rm Tr}\left[ \hbar ^{2}[{\bf 
H}_{S},[{\bf H}_{S},\tilde{\rho}(0)]]\tilde{\rho}(0)+[{\bf H}_{S},\tilde{\rho
}(0)]{\tt L}^{\prime }(t)[\tilde{\rho}(0)]_{t=0}+\tilde{\rho}(0)\left\{ 
\frac{\partial {\tt L}^{\prime }(t)}{\partial t}[\tilde{\rho}(0)]\right\}
_{t=0}\right] ,
\end{equation}
where ${\tt L}^{\prime }$ is now given by Eq.~(\ref{eq:newterms}). As in the
semigroup analysis above, the second of these traces does not vanish (nor
does the first, but it will be cancelled due to a contribution from the
third).

Thus we find that for the {\it dynamical} fidelity, the effect of a symmetry
breaking perturbation results in a second order time instability in the
system for both Markovian and non-Markovian situations. It is remarkable
that while the memory fidelity is {\em completely} robust to first order ($
\epsilon $) perturbing errors, the dynamical fidelity does not show the same
robustness, with instability arising at second order in time. In terms of
real quantum computation therefore, this implies that DF subspaces must be
supplemented by QECC in order to be truly useful beyond merely providing
high quality quantum memory. In particular, the instability in the dynamical
fidelity implies that in order to realize the robustness of DF subspaces to
symmetry breaking perturbations, operations on the DF subspace must be
performed over a time scale short in comparison with the perturbing error
rate. Thus, for the realistic scheme in which DF subspaces are supplemented
by quantum error correcting codes (by, for example, concatenating the DF
subspaces within QECC as in Ref.\cite{Lidar:98QECC-DFS}), if the operations
performed on the DF subspace in order to execute the QECC are executed
frequently and rapidly, the full scheme can provide a significant
improvement over pure DF subspaces.

\section{Summary and Conclusions}

\label{conc}

We have shown here how the formally exact operator sum representation (OSR)
for the time evolution of the density matrix can be cast in a form which
bears a significant resemblance to the semigroup master equation (SME),
through the introduction of a time-independent (fixed) operator basis. By
using this fixed-basis OSR equation, we were able to easily calculate the
fidelity in the OSR, as well as to provide a derivation of the SME which
makes explicit the role played by the coarse-graining assumption. Somewhat
surprisingly, we found an important difference between the OSR and the SME,
namely, the first order decoherence rate {\em always} vanishes in the
former, but not always in the latter, for a finite Hamiltonian. This effect
is readily traced to the coarse-graining time averaging assumption within
our derivation. This result is significant for both error-correction schemes
aimed at improving the fidelity, and for commonly used simplified models of
decoherence. We illustrated the latter with the well-known case of
phase-damping.

Using the fixed-basis OSR, we have then undertaken a detailed study of
short-time expansions of the mixed-state fidelity under a variety of
conditions. The mixed-state fidelity provides a measure of the extent of
quantum coherence in the system. We have examined both the usual mixed-state
memory fidelity relevant to quantum memory, and a dynamical fidelity which
we defined to act as a measure of coherence for a time-evolving system. Our
main achievement here has been to extend the robustness results of Ref.~\cite
{Lidar:PRL98} regarding decoherence-free (DF) subspaces. For the
preservation of quantum {\em memory}, we showed that in both the OSR and SME
approaches, DF subspaces are stable to {\em all orders in time} to a
symmetry-breaking perturbation. The first errors entering the quantum memory
can therefore only be $O(\epsilon ^{2})$, where $\epsilon $ measures the
strength of the perturbation. This result goes beyond the first order
stability result of $O(\epsilon t^{2})$ arrived at in Ref.~\cite{Lidar:PRL98}
, which was restricted to the Markovian case. It shows that {\em 
DF subspaces are indeed ideal for quantum memory in all situations}.
In making this statement, we note that we have not shown here how to
perform input and output to the DF subspaces.  Further investigation
is needed to address this issue.  The quantum memory stability results
are summarized in Table \ref{datable}.  For the dynamical fidelity, a
weaker result is obtained, namely that this has only a vanishing {\em
first} order decoherence rate under a perturbation. Thus the first
errors entering the dynamical fidelity can be $O(\epsilon t^{2})$.  

This stability analysis of static and dynamic fidelities is of particular
relevance for practical implementations of quantum computation. The complete
stability to perturbations of static fidelity within DF subspaces is very
encouraging for use in quantum memory.  Thus passive error correction appears
to be sufficient for this. In contrast, the weaker first order stability
condition derived for the dynamical fidelity within DF subspaces implies
that application of active error-correction techniques will likely be
necessary in order to preserve coherence {\em during} computation. Further,
if the operations performed on the DF subspace in order to execute active
error-correction are executed rapidly, then the full DF stability to
symmetry breaking perturbations can be recovered. Such a scheme of combining
active and passive error-correction is possible by concatenating codes
constructed from DF subspaces with active quantum error correction, as
demonstrated in Ref.~\cite{Lidar:98QECC-DFS}.

\section*{Acknowledgments}

This material is based upon work supported by the U.S. Army Research Office
under contract/grant number DAAG55-98-1-0371. We would like to thank Dr. I.
Chuang for helpful discussions and Dr. P. Zanardi for useful comments. Part
of this work was completed during the Elsag-Bailey -- I.S.I. Foundation
research meeting on quantum computation in 1998.

\begin{table}[tbp]
\caption{Different order decoherence rates for the mixed-state memory ($
\protect\tau $) and dynamical ($\bar{\protect\tau}$) fidelities under the
various conditions cited in the text. Unless otherwise noted, $n\geq 1$.}
\label{datable}
\begin{tabular}{lcc}
& SME & OSR \\ 
\tableline General & $1/\tau_{1}\neq 0$ & $1/\tau_{1}=0 \tablenotemark[1]$
\\ 
& $1/\tau_{n}\neq 0$, $n\geq 2$ & $1/\tau_{n}\neq 0$, $n\geq 2$  \\ 
DF subspaces & $1/\tau_{n}=0 \tablenotemark[2]$ & $1/\tau_{n}=0$ \\ 
memory fidelity for $\epsilon $-perturbed DF subspaces & $1/\tau_n=0$ & $
1/\tau_n=0$ \\ 
dynamical fidelity for $\epsilon$-perturbed DF subspaces & $1/\bar{\tau}
_{1}=0$ & $1/\bar{\tau}_{1}=0$ \\ 
& $1/\bar{\tau}_{n}\neq 0$, $n\geq 2$  & $1/\bar{\tau}_{n}\neq 0$, $n\geq 2$ 
\end{tabular}
\tablenotemark[1]{\small {finite bath and finite total Hamiltonian only.}} \newline
\tablenotemark[2]{\small {both with and without perturbing Lamb shift.}} \newline
\end{table}


\begin{thebibliography}{10}
\bibitem{zurek}
An interesting example of this general first-order stability is provided by the 
recently analysed situation of environmentally induced superselection of
energy eigenstates for a Brownian particle interacting with an adiabatically 
slow bath of harmonic oscillators~\cite{Paz:99}.
\end{thebibliography}
\end{document}